\documentclass[superscriptaddress,twocolumn,showpacs,longbibliography,
amssymb,amsmath,nobibnotes,aps,prd,
nofootinbib]{revtex4-1}
\usepackage{graphicx,subfigure,bm,color,psfrag,hyperref}
\usepackage{float}
\usepackage{epsf}
\usepackage{bm}
\usepackage{booktabs}
\usepackage{amsmath}
\usepackage{amsfonts}
\usepackage{amssymb}
\usepackage{epstopdf}
\usepackage{natbib}
\usepackage[normalem]{ulem}
\usepackage[dvipsnames]{xcolor}
\usepackage{array}
\providecommand{\U}[1]{\protect\rule{.1in}{.1in}}
\newcommand{\be}{\begin{equation}}
\newcommand{\ee}{\end{equation}}

\newcommand{\mincir}{\raise
-3.truept\hbox{\rlap{\hbox{$\sim$}}\raise4.truept\hbox{$<$}\ }}
\newcommand{\magcir}{\raise
-3.truept\hbox{\rlap{\hbox{$\sim$}}\raise4.truept\hbox{$>$}\ }}

\newtheorem{remark}{Remark}[section]

\definecolor{blue}{rgb}{0.36, 0.54, 0.66}
\definecolor{amaranth}{rgb}{0.9, 0.17, 0.31}
\definecolor{pink}{rgb}{0.87, 0.56, 0.81}
\definecolor{ao}{rgb}{0.0, 0.5, 0.0}
\definecolor{maroon}{rgb}{0.76, 0.13, 0.28}
\definecolor{cardinal}{rgb}{0.77, 0.12, 0.23}
\definecolor{lightcardinal}{rgb}{0.97, 0.42, 0.53}
\definecolor{frenchlila}{rgb}{0.53, 0.38, 0.56}
\definecolor{yellow}{rgb}{1.0, 1.0, 0.87}
\definecolor{lightseagreen}{rgb}{0.7, 0.92, 0.68}
\definecolor{gray}{rgb}{0.9, 0.9, 0.9}
\definecolor{lightblue}{rgb}{0.66, 0.84, 0.96}

\hypersetup{colorlinks,linkcolor={maroon},citecolor={magenta},urlcolor={ao}}

\begin{document}

\title{Gravitational reheating 
 formulas and bounds in oscillating backgrounds II: Constraints on the spectral index and gravitational dark matter production}

\author{Jaume de Haro}
\email{jaime.haro@upc.edu}
\affiliation{Departament de Matem\`atiques, Universitat Polit\`ecnica de Catalunya, Diagonal 647, 08028 Barcelona, Spain}

\author{Supriya Pan}
\email{supriya.maths@presiuniv.ac.in}
\affiliation{Department of Mathematics, Presidency University, 86/1 College Street,  Kolkata 700073, India}
\affiliation{Institute of Systems Science, Durban University of Technology, PO Box 1334, Durban 4000, Republic of South Africa}

\begin{abstract}

The reheating temperature plays a crucial role in the early universe's evolution, marking the transition from inflation to the radiation-dominated era. It directly impacts the number of 
$e$-folds and, consequently, the observable parameters of inflation, such as the spectral index of scalar perturbations. By establishing a relationship between the gravitational reheating temperature and the spectral index, we can derive constraints on inflationary models. Specifically, the range of viable reheating temperatures imposes bounds on the spectral index, which can then be compared with observational data, such as those from the Planck satellite, to test the consistency of various models with cosmological observations. Additionally, 
in the context of dark matter production, we demonstrate that gravitational reheating provides a viable mechanism when there is a relationship between the mass of the dark matter particles and the mass of the particles responsible for reheating. This connection offers a pathway to link dark matter genesis with inflationary and reheating parameters, allowing for a unified perspective on early universe dynamics.

\end{abstract}

\vspace{0.5cm}


\maketitle

\thispagestyle{empty}

\section{Introduction}

The relation between the reheating temperature and the spectral index of scalar perturbations plays a crucial role in understanding the dynamics of the early universe, particularly in the context of inflationary cosmology~\cite{Guth:1980zm,Linde:1981mu}. Inflation, a rapid expansion of the universe in its earliest moments, produces scalar perturbations, which later evolve into the large-scale structures we observe today, see for instance Refs.~\cite{Barrow:1981pa,Spokoiny:1984bd,Steinhardt:1984jj,Lucchin:1984yf,Hawking:1984kj,Lyth:1984gv,Belinsky:1985zd,Mijic:1986iv,Khalfin:1986ui,Silk:1986vc,Mijic:1986iv,Burd:1988ss,Olive:1989nu,Ford:1989me,Adams:1990pn,Freese:1990rb,Wang:1991ww,Polarski:1992dq,Liddle:1992wi,Linde:1993cn,Barrow:1993hn,Barrow:1994nt,Vilenkin:1994pv,Peter:1994dx,Sasaki:1995aw,Barrow:1995xb,Lidsey:1995np,Parsons:1995ew,Liddle:1998wc,Guth:2000ka,Riotto:2002yw,Feinstein:2002aj,Ashcroft:2002vj,Boubekeur:2005zm,Conlon:2005jm,Ferraro:2006jd,Cheung:2007st,Chen:2008wn,Baumann:2008bn,Koivisto:2008xf,Pal:2009sd,Martin:2013nzq,Martin:2013tda,Sebastiani:2013eqa,Hamada:2014iga,Freese:2014nla,vandeBruck:2015xpa,vandeBruck:2015gjd,vandeBruck:2016rfv,Ratra:2017ezv,Rubio:2018ogq,Giare:2019snj,vandeBruck:2021xkm,Forconi:2021que,Odintsov:2023weg,Giare:2023wzl,Jinno:2023bpc,Garfinkle:2023vzf}. The spectral index, 
$n_s$, characterizes the distribution of these perturbations and offers insights into the physics of the inflationary period.

In addition, 
the number of $e$-folds between the 
horizon crossing and 
the end of inflation  is influenced by the reheating temperature, $T_{\rm reh}$. Since the spectral index 
$n_s$ depends on this last number of $e$-folds, the reheating temperature impacts the predicted value of 
$n_s$. Higher reheating temperatures generally reduce the duration of the reheating phase, leading to fewer $e$-folds and a lower predicted value of the spectral index, while lower reheating temperatures result in more $e$-folds and a higher spectral index.
As a consequence, 
by studying the connection between  $T_{\rm reh}$ and $n_s$,  different inflationary models can be tested
   and compared with the observational data, such as the measurements from the Planck satellite \cite{Planck:2018jri}. This relationship serves as a valuable tool for constraining models of inflation, 
   and  only those predicting a viable range of $n_s$
  consistent with observations, are considered plausible.

In the context of gravitational reheating and considering a heavy scalar quantum field conformally coupled to gravity, the created particles must decay into lighter ones to reheat the universe. First, continuing with our previous work \cite{deHaro:2024xgd},  we derive the formula for the reheating temperature as a function of the decay rate and the mass $m_{\chi}$
  of the produced particles, identifying viable values of 
$m_{\chi}$
  and the range of possible reheating temperatures. Using this range and considering the relationship between the reheating temperature and the spectral index, we are able to constrain the latter. Indeed, these spectral index values are more tightly constrained than those obtained experimentally by Planck’s data \cite{Planck:2018jri}.

On the other hand, 
in the absence of strong couplings between the inflaton and standard particles, gravitational reheating can produce a non-thermal spectrum of particles, including potential dark matter candidates. This mechanism is especially intriguing in models where dark matter particles have minimal interactions with ordinary matter. During gravitational reheating, dark matter production can occur efficiently through gravitational interactions alone, providing a natural and model-independent production pathway. This scenario could help explain the relic density of dark matter observed today without requiring direct couplings to the inflaton field.
Having in mind this fact,  we investigate the gravitational production of dark matter within the framework of gravitational reheating. Specifically, we consider two distinct scalar fields conformally coupled to gravity, which generate two types of particles with different masses. One type decays into lighter particles, facilitating the reheating of the universe, while the other, a candidate for explaining the dark matter content, does not decay. In this work, we determine the range of particle masses that yield both a viable reheating temperature and a present-day value for the cold dark matter energy density, thereby contributing to a more complete understanding of the dark matter and energy composition of the universe.

The paper is structured as follows: In Section~\ref{sec-II}, we use the Wentzel-Kramers-Brillouin (WKB) method in the complex plane to analytically calculate the $\beta$-Bogoliubov coefficient, the key component for determining particle production. In Section~\ref{sec-III}, we constrain the viable values of the spectral index by relating it to the reheating temperature through the last few $e$-folds. In Section~\ref{sec-IV-application}, we derive an analytical formula for the reheating temperature when reheating occurs via gravitational particle production. This formula allows us to identify the viable mass range for the produced particles and further constrain the spectral index. Section~\ref{sec-V} addresses the gravitational production of dark matter, where we establish the relationship between the mass of dark matter and that of the scalar field responsible for gravitational reheating. Finally, in Section~\ref{sec-summary} we summarize the present work.

Throughout  the manuscript we  use natural units, i.e., 
 $\hbar=c=k_B=1$,  and the reduced Planck's mass is denoted by $M_{\rm pl}\equiv \frac{1}{\sqrt{8\pi G}}\cong 2.44\times 10^{18}$ GeV. 

\section{Analytic calculation of the energy density of produced particles: The Stokes phenomenon}
\label{sec-II}

We consider gravitational reheating through the production of heavy massive particles generated by a spectator scalar field conformally coupled to the Ricci scalar.
 After their creation, these particles decay into Standard Model particles, which, after thermalization, eventually dominate the inflaton's energy density, thereby reheating the universe.

The key quantity to calculate is the energy density of the particles produced at the end of inflation \cite{Zeldovich:1971mw}:

\begin{eqnarray}
    \langle \rho_{\rm END}\rangle =\frac{m_{\chi}}{2\pi^2}
\int_0^{\infty}k^2|\beta_k|^2dk,
\end{eqnarray}
where $\beta_k$ is the $\beta$-Bogoliubov coefficient \cite{Bogolyubov:1958km}
and  $m_{\chi}$ is the mass of the produced particles, which we will assume to be small compared to the Hubble rate at the end of inflation $H_{\rm END}$.

  We focus on the following class of potentials
  studied in \cite{Drewes:2017fmn} (although our analysis can be applied to inflationary models such as
  Hyperbolic Inflation (HBI), Superconformal $\alpha$-Attractor B Inflation (SABI) or Superconformal $\alpha$-Attractor T Inflation (SATI) \cite{Martin:2013tda})
  which allow for efficient gravitational reheating while maintaining consistency with observational data on the early universe:
\begin{eqnarray}\label{potential}
    V_n(\varphi)=\lambda M_{\rm pl}^4\left(1- e^{-\sqrt{\frac{2}{3}}\frac{\varphi}{M_{\rm pl}}}\right)^{2n}\quad\mbox{with}\qquad n>2, 
\end{eqnarray}
where {$n$ is a natural number and} the value of the dimensionless constant $\lambda \sim 3\pi^2(1-n_s)^2 10^{-9}$ in which $n_s\sim 0.96$ is the value of the spectral index of scalar perturbations, is obtained from the power spectrum of scalar perturbations.
 During oscillations, since close to the minimum, the potential behaves like $\varphi^{2n}$,  using the virial theorem we find that the effective Equation of State (EoS) parameter is $w_{\rm eff}(n)=\frac{n-1}{n+1}$ \cite{Turner:1983he}, which is higher than $1/3$ for $n>2$, which ensures that the energy density of the  inflaton decreases faster than the energy density of the produced particles and their decay products, and thus, this last one eventually will dominate leading to a successful reheating of the universe.

\begin{remark}
 One can also make viable the cases $n=1$ and $n=2$ assuming that the inflaton decays into 
    relativistic particles of the Standard Model (SM)~\cite{Garcia:2020wiy, Kaneta:2022gug}. However, this means that the  quantum field is coupled to the inflaton field, and thus, this is not exactly what one understand as a ``gravitational particle production''~\cite{Ford:1986sy, Chun:2009yu, Lankinen:2016ile,Haro:2018zdb}, where the quantum field is coupled to the Ricci scalar. On the other hand, one can consider that effectively, the reheating occurs due to the coupling with the inflaton field and a quantum field producing the particles of the SM, and also considers another spectator massive quantum field  coupled to gravity, which would be a candidate to produce gravitationally dark matter ~\cite{Jenks:2024fiu, Ema:2018ucl, Chung:2001cb, Garcia:2020eof}.
\end{remark}

Taking into account that when $m_{\chi} \ll H_{\rm END}$, the main contribution of particle production comes from the pure Hubble expansion~\cite{Hashiba:2018iff}, we disregard the oscillating effect, and we approximate the  
 scale factor, in the complex plane,   by:
 \begin{widetext}
\begin{eqnarray}
a(\eta)=\left\{\begin{array}{ccc}
       -\frac{1}{H_{\rm END}\eta}  &\mbox{when}& \frak{Re}(\eta)<\eta_{\rm END}\\ 
       & &\\
       a_{\rm END}\left[\frac{3n}{n+1}\left(1-\frac{\eta}{\eta_{\rm END}}\right)+ \frac{\eta}{\eta_{\rm END}}
       \right]^{\frac{n+1}{2n-1}}
         &\mbox{when}& 
         \frak{Re}(\eta)\geq \eta_{\rm END},
    \end{array}\right.
\end{eqnarray}
\end{widetext}
with $a_{\rm END}=-\frac{1}{H_{\rm END}\eta_{\rm END}}$, being $\eta_{\rm END}<0$.
This model depicts a purely de Sitter phase during inflation 
with a sudden transition to a phase with a constant EoS parameter $w_{\rm eff}(n)=\frac{n-1}{n+1}$. We have to recall that, for a smooth potential as (\ref{potential}), this phase transition is not so abrupt.

On the other hand, 
the complex WKB approximation tells us that
 the main contribution of the particle production comes from the turning points (Stokes phenomenon) --- in our case points where $\omega_k^2(\eta)=k^2+a^2(\eta)m_{\chi}^2=0$ --- and at the end of inflation \cite{Hashiba:2021npn}. 

\subsection{Turning points}

The turning points corresponding to the de Sitter phase are
$\eta_c=\pm i \frac{m_{\chi}}{H_{\rm END}k}$, and since its real part is $0$ they do not belong to the inflationary domain, and thus, they do not contribute to the particle production. The other turning points, corresponding to the other phase,  are: 
\begin{align}
    \eta_c=\frac{3n}{2n-1}\eta_{\rm END}-(-1)^{\frac{2n-1}{2(n+1)}} \left(\frac{n+1}{2n-1} \right)\nonumber\\ \times \eta_{\rm END}\left( \frac{k}{a_{\rm END}m_{\chi}}\right)^{\frac{2n-1}{n+1}}.
\end{align}
Here we can see that for $n=2$, the real part of the turning points is $2\eta_{\rm END}$, which belongs to the inflationary domain, and thus, there is no contribution to the particle production. For the other values of $n$, when  $k\gg a_{\rm END}m_{\chi}$, we can see that there are turning points whose real part does not belong to the inflationary domain, and thus, they contribute to the particle production.

Next, we consider only the turning points with positive imaginary part and we 
take the straight line: 
\begin{align}\label{line}
\eta=\frac{3n}{2n-1}\eta_{\rm END}-(-1)^{\frac{2n-1}{2(n+1)}}\left(\frac{n+1}{2n-1}\right) \nonumber\\ \times \eta_{\rm END}\left( \frac{k}{a_{\rm END}m_{\chi}}\right)^{\frac{2n-1}{n+1}}\tau, \quad\mbox{with}
\quad 0\leq \tau\leq 1.
\end{align}
Over this line $a(\tau)=i
\frac{k}{m_{\chi}}\tau^{\frac{n+1}{2n-1}}
$, and thus, $\omega_k^2(\tau)
=k^2(1- \tau^{\frac{2(n+1)}{2n-1}})$. 
Therefore, based on the approach described in  \cite{landau1991quantum},  we have:
\begin{eqnarray}\label{Im}
    |\beta_k|^2
    \cong \exp\left(-4\frak{Im}
\int_{
\frac{3n}{2n-1}\eta_{\rm END}}^{\eta_c}\omega_k(z) dz
    \right),
\end{eqnarray}
    where $\frak{Im}$ denotes the imaginary part and the integral is performed along the line (\ref{line}).
    Then,  
we have:
\begin{align}
    |\beta_k|^2
    \cong
    \exp\left(-c_n\left(
    \frac{k}{H_{\rm END}a_{\rm EDN}}\left( \frac{H_{\rm END}}{m_{\chi}}\right)^{\frac{2n-1}{3n}}
    \right)^{\frac{3n}{n+1}}
    \right),
\end{align}
where the value of the  dimensionless constant is:
\begin{eqnarray}
&&c_n=4
\sin\left(\frac{\pi}{2(n+1)}\right)
    \frac{n+1}{2n-1}
    \int_0^1\sqrt{1-\tau^{\frac{2(n+1)}{2n-1}}}d\tau
\nonumber\\ &&=2\sin\left(\frac{\pi}{2(n+1)}\right)  
\int_0^1 t^{-\frac{3}{2n+2}} \sqrt{1-t}dt\nonumber\\
&&=2\sin\left(\frac{\pi}{2(n+1)}\right)
B
\left(\frac{2n-1}{2n+2},\frac{3}{2}\right),   
\end{eqnarray}
where  $B$ denotes the Euler's beta function, and
from all the turning points 
with positive imaginary part we have chosen the one
which the minimum value of the imaginary part  in the exponential of (\ref{Im})
(see for details Section $52$ of  Chapter VII of \cite{landau1991quantum}). 

Finally,  its contribution to the energy density is:
\begin{eqnarray}
C_n m_{\chi}H_{\rm END}^3
    \left(\frac{m_{\chi}}{H_{\rm END}} \right)^{(2n-1)/n}\left(\frac{a_{\rm END}}{a(t)}\right)^3,
\end{eqnarray}
where
\begin{eqnarray}
&&C_n \cong \frac{1}{2\pi^2}
\int_{\left(\frac{m_{\chi}}{H_{\rm END}}\right)^{\frac{n+1}{3n}}}^{\infty}x^2
\exp\left({-c_nx^{\frac{3n}{n+1}}}\right)dx \nonumber\\ &&\cong 
\frac{1}{2\pi^2}
\int_0^{\infty}x^2
\exp\left({-c_nx^{\frac{3n}{n+1}}}\right)dx.
\end{eqnarray}

\subsubsection{Case $n=1$}

As an example, we calculate:
\begin{align}
c_1=4\sqrt{2}\int_0^1\sqrt{1-\tau^4}d\tau=\sqrt{2}B
    \left(\frac{1}{4},\frac{3}{2}\right)=\sqrt{2}\frac{
    \Gamma\left(\frac{1}{4}\right)
    \Gamma\left(\frac{3}{2}\right)}{
    \Gamma\left(\frac{7}{4}\right)},
\end{align}
and since 
$\Gamma\left(\frac{3}{2}\right)=\frac{\sqrt{\pi}}{2}$,
and using the duplication formula \cite{abramowitz1965handbook}
\begin{eqnarray}
\Gamma(z)\Gamma\left(z+\frac{1}{2} \right)=2^{1-2z}\sqrt{\pi}
\Gamma(2z),
\end{eqnarray}
we find $\Gamma\left(\frac{7}{4}\right)=\frac{3\pi}{2\sqrt{2}}\Gamma^{-1}\left(
\frac{1}{4}\right)$, and thus, we obtain the same result numerically obtained in \cite{Hashiba:2018iff} and analytically in the formula (B.4) of \cite{Jenks:2024fiu}:
\begin{eqnarray}
    c_1=\frac{2}{3\sqrt{\pi}}
    \Gamma^{2}\left(
\frac{1}{4}\right)\cong 4.9442,\end{eqnarray}
where we have used that 
$\Gamma\left(
\frac{1}{4}\right)\cong 3.6256$.  Having the value of $c_1$, one now has:
\begin{eqnarray}
&& C_1\cong \frac{1}{2\pi^2}
    \int_0^{\infty}x^2e^{-c_1x^{3/2}}dx=\frac{1}{\pi^2c_1^2}
\int_0^{\infty}y^5e^{-y^2}dy \nonumber\\  
&& =
\frac{1}{\pi^2c_1^2}=
\frac{9}{4\pi}\Gamma^{-4}\left(\frac{1}{4}\right)
\cong 4\times 10^{-3}.   
\end{eqnarray}

\subsection{End of inflation}

Concerning to  the end of inflation, $\eta_{\rm END}$, close to this point we make the quadratic approximation:
\begin{eqnarray}
a(\eta)\cong a_{END}+a_{END}^2H_{END}(\eta-\eta_{END}) \nonumber\\+\frac{1}{2}a_{END}^3H_{END}^2(\eta-\eta_{END})^2,
\end{eqnarray}
and up to degree two, we have
\begin{eqnarray}\label{aquadrat}
a^2(\eta)\cong a_{END}^2+2a_{END}^3H_{END} (\eta-\eta_{END}) \nonumber\\+{2}a_{END}^4H_{END}^2 (\eta-\eta_{END})^2 ,
\end{eqnarray}
and the frequency can be approximated by
\begin{eqnarray}
&& \omega_k^2(\eta)
\cong  k^2 +\frac{m_{\chi}^2a_{END}^2}{2}
\nonumber\\ &&+2
a_{END}^4H_{END}^2m_{\chi}^2
\left(\eta-\frac{3}{2}\eta_{END} \right)^2,\end{eqnarray}
{and defining 
\begin{eqnarray}
\tau\equiv\sqrt{\sqrt{2} a^2_{\rm END} H_{\rm END} m_{\chi}}
\left(\eta-\frac{3}{2}\eta_{\rm END}\right),
\end{eqnarray}
the dynamical equation of the $k$-mode is
\begin{eqnarray}\label{kg2}
\frac{d^2\chi_k}{d\tau^2}+(\kappa^2+\tau^2)\chi_k=0,
\end{eqnarray}
where we have introduced the notation 
$\kappa^2=\frac{k^2+{ a^2_{\rm END}
m_{\chi}^2}/2}{\sqrt{2} a^2_{\rm END} H_{\rm END} m_{\chi}
 }$.
Note that for this quadratic frequency the $\beta$-Bogoliubov coefficient is obtained using the well-known formula 
{
\begin{eqnarray}\label{beta2}
&& |\beta_k|^2= e^{-\pi\kappa^2}=
       \exp\left(-\pi\frac{k^2+{ a^2_{\rm END}
       m_{\chi}^2}/2}{\sqrt{2} a^2_{\rm END} H_{\rm END} m_{\chi}  }\right)
       \nonumber\\
&& \cong 
       \exp\left(-\pi\frac{k^2}{\sqrt{2} a^2_{\rm END} H_{\rm END} m_{\chi}  }\right)       ,
   \end{eqnarray}
  where we have used that $m_{\chi}\ll H_{\rm END}$, and which practically  coincides 
(the difference is the substitution, in the exponential, of $\pi$ by $4$) with the formula (41) obtained in \cite{Chung:1998bt} using the steepest descent method. }}
This result can be derived as follows.
First, recall  that the positive frequency modes in the WKB approximation are
\begin{eqnarray}
\phi_{k,+}(\tau)=\frac{1}{(\kappa^2+\tau^2)^{1/4}}e^{-i\int \sqrt{\kappa^2+\tau^2}d\tau},
\end{eqnarray}
and for large values of $|\tau|$ ($|\tau|\gg \kappa$),  one can make the approximations
$(\kappa^2+\tau^2)^{1/4}\cong |\tau|^{1/2}$ and $\sqrt{\kappa^2+\tau^2}\cong |\tau|\left(1+\frac{\kappa^2}{2\tau^2}  \right)$, obtaining

\begin{eqnarray}
\left\{\begin{array}{ccc}
       \phi_{k,+}(\tau\ll -\kappa)\cong |\tau|^{-1/2+i\kappa^2/2}e^{i \tau^2/2},\\
       \bigskip \\
       \phi_{k,+}(\tau\gg \kappa)\cong |\tau|^{-1/2-i\kappa^2/2}e^{-i\tau^2/2},
    \end{array}\right.
\end{eqnarray}
while for the negative frequency modes 

\begin{eqnarray}
&&\phi_{k,-}(\tau\gg \kappa)=\frac{1}{(\kappa^2+\tau^2)^{1/4}}e^{i\int \sqrt{\kappa^2+\tau^2}d\tau}\nonumber\\ 
&& \cong |\tau|^{-1/2+i\kappa^2/2}e^{i \tau^2/2}.
\end{eqnarray}
On the other hand,  the positive frequency  modes evolve as 
\begin{align}\label{evolution}
\phi_{k,+}(\tau\ll -\kappa)\longrightarrow \alpha_k\phi_{k,+}(\tau\gg \kappa)+\beta_k\phi_{k,-}(\tau\gg \kappa).
\end{align}
Thus, in order to calculate the Bogoliubov coefficients one can also use the WKB method in the complex plane integrating the frequency along the path $\gamma=\{ z=|\tau|e^{i\alpha}, -\pi \leq \alpha\leq 0\}$, obtaining that for $\tau\gg \kappa$, the early time positive frequency modes evolve at late time as
\begin{eqnarray}
e^{-\frac{\kappa^2}{2}\pi}|\tau|^{-1/2+i\kappa^2/2}e^{i\tau^2/2},
\end{eqnarray}
and comparing with (\ref{evolution}) one gets

\begin{align}
|\beta_k|^2 \cong e^{-\kappa^2\pi},~~ \mbox{and}~~ |\alpha_k|^2=1+|\beta_k|^2=1+e^{-\kappa^2\pi}.
\end{align}
Note that we can obtain the same result considering the turning point
\begin{eqnarray}
    \eta_c=\frac{3}{2}\eta_{\rm END}+ i \frac{\sqrt{k^2+{ a^2_{\rm END}
m_{\chi}^2}/2}}{\sqrt{2} a^2_{\rm END} H_{\rm END} m_{\chi}
 },
\end{eqnarray}
and calculating $4\frak{Im}
\int_{\frac{3}{2}\eta_{\rm END} }^{\eta_c}\omega_k(z)dz$
over the line
$
\frac{3}{2}\eta_{\rm END}+ i \frac{\sqrt{k^2+{ a^2_{\rm END}
m_{\chi}^2}/2}}{\sqrt{2} a^2_{\rm END} H_{\rm END} m_{\chi}
 }\tau
$
with $0\leq \tau\leq 1$, 
one gets
$\pi \kappa^2$. Then, the complex WKB method also leads to the same result.

From  this result, and considering the case $m_{\chi}\ll H_{\rm END}$, we obtain that the contribution to the energy density of the produced particles,  is:
\begin{eqnarray}
    \frac{1}{4\pi^3}m_{\chi}^2H_{\rm
    END}^2
    \sqrt{\frac{m_{\chi}}{\sqrt{2}H_{\rm END}}}
    \left(\frac{a_{\rm END}}{a(t)}\right)^3.
\end{eqnarray}

\subsection{Energy density of the produced particles at the end of inflation}

Taking into account the contribution of the turning points and  the end of inflation,  we can conclude that, for $m_{\chi}\ll H_{\rm END}$, the energy density of the scalar produced particles at the end of inflation is:
\begin{eqnarray}
    \langle \rho_{\rm END}\rangle\cong
    \left\{\begin{array}{ccc}  \frac{9}{4\pi}\Gamma^{-4}\left(\frac{1}{4}\right)
m_{\chi}^2H^2_{\rm END}    & \mbox{for}  &  n=1 \\
   & &\\
\frac{1}{4\pi^3}m_{\chi}^2H^2_{\rm END}\sqrt{\frac{m_{\chi}}{\sqrt{2}H_{\rm END}}}& \mbox{for}& n\not=1,
    \end{array}\right.
\end{eqnarray}
where we have used that, for $n\not=1$, the main contribution to the particle production is at the end of inflation, because, since we are dealing with particles with mass lower than the Hubble rate at the end of inflation, 
the contribution 
$ C_n
    m_{\chi}H_{\rm END}^3
    \left(m_{\chi}/H_{\rm END} \right)^{(2n-1)/n}$ decreases as $n$ increases, so the dominant term is for $n=2$ which is of the same order as the contribution 
    produced close to  $\eta_{\rm END}$.
{
}

\section{Relationship between the reheating temperature and the spectral index}
\label{sec-III}

For the family of potentials (\ref{potential}), 
the main slow roll parameter $\epsilon=\frac{M_{\rm pl}^2}{2}\left(\frac{\partial_{\varphi}V_n}{V_n} \right)^2$, is given by:
\begin{align}
    \epsilon=\frac{4n^2}{3}\frac{
e^{-2\sqrt{\frac{2}{3}}\frac{\varphi}{M_{\rm pl}}}    
    }{
\left(1- e^{-\sqrt{\frac{2}{3}}\frac{\varphi}{M_{\rm pl}}}\right)^{2}},
\end{align}
and at the horizon crossing, where $
e^{-\sqrt{\frac{2}{3}}\frac{\varphi_*}{M_{\rm pl}}}\ll 1$, can be related to the spectral index of scalar perturbations, 
$n_s$,  as follows:
\begin{eqnarray}
 H_*\cong \sqrt{10}\pi (1-n_s) 10^{-5}M_{\rm pl}, \quad
 \epsilon_*=\frac{3}{16}(1-n_s)^2,
\end{eqnarray}
where the star $(*)$ denotes that the quantities are evaluated at the horizon crossing.  On the other hand, at the end of inflation, which occurs when $\epsilon=1$, one can write the Hubble rate  as a function of the spectral index of scalar perturbations, as follows:
\begin{align}\label{END}
&& \varphi_{\rm END}=-\sqrt{\frac{3}{2}}\ln\left(\frac{2\sqrt{3}n-3}{4n^2-3} \right)M_{\rm pl}
\Longrightarrow \nonumber\\
&& H_{\rm END}^2=
\frac{3\pi^2}{2}(1-n_s)^2
    \left(1- \frac{2\sqrt{3}n-3}{4n^2-3}    \right)^{2n} 10^{-9}M_{\rm pl}^2,
\end{align}
where we have used that at the end of inflation one has:
\begin{eqnarray}
    3M_{\rm pl}^2H_{\rm END}^2
    =\frac{3}{2}V(\varphi_{\rm END}).
\end{eqnarray}

\subsection{The last number of 
$e$-folds}
The most useful formula that relates the last number of $e$-folds (from the horizon crossing to the end of inflation) with the reheating temperature $T_{\rm reh}(n)$, is given by~\cite{Rehagen:2015zma}:
\begin{align}\label{efols}
&& N_*(n)\cong 54.2298+\frac{1}{2}\ln \epsilon_*+
\frac{1}{3(1+w_{\rm eff}(n))}
\ln\left(
\frac{
M_{\rm pl}^2}{3H_{\rm END}^2}\right) \nonumber\\ 
 && +\frac{3w_{\rm eff}(n)-1}{12(1+w_{\rm eff}(n))}\ln\left(
    \frac{30M_{\rm pl}^4}{g_{\rm reh}\pi^2T_{\rm reh}^4(n)}
    \right),
\end{align}
where  inserting the corresponding expressions, we get:
\begin{eqnarray}\label{efols2}
&& N_*(n)\cong 53.3928+
\frac{2n-1}{3n}\ln(1-n_s)
\nonumber\\ && -\frac{n+1}{3}\ln
\left(\frac{4n^2-2\sqrt{3}n}{4n^2-3}
\right)
+\frac{n+1}{6n}\ln\left(
\frac{2\times 10^9}{9\pi^2}
\right)
\nonumber\\ &&+\frac{n-2}{12n}\ln\left(
    \frac{30M_{\rm pl}^4}{g_{\rm reh}\pi^2T_{\rm reh}^4(n)}
    \right).
\end{eqnarray}
The important point is that we are considering models with $w_{\rm eff}(n)>1/3$, which means that the last term in (\ref{efols}) is a decreasing function with respect to the reheating temperature. Thus, the higher the reheating temperature, the greater the last number of $e$-folds.

Next, we  calculate the last number of $e$-folds using the slow roll approximation, as follows:
\begin{eqnarray}\label{n-slow-roll}
&& N_*(n)=\frac{1}{M_{\rm pl}}\int_{\varphi_{\rm END}}^{\varphi_*}\frac{1}{\sqrt{2\epsilon}}d\varphi
    \nonumber\\ && =\frac{\sqrt{3}}{2\sqrt{2}nM_{\rm pl}}\int_{\varphi_{\rm END}}^{\varphi_*}
\left(e^{\sqrt{\frac{2}{3}}\frac{\varphi}{M_{\rm pl}}}-1
    \right)
    d\varphi\nonumber\\
   && = \frac{2}{1-n_s}-\frac{3}{4n}
    \ln\left(\frac{8n}{3(1-n_s)} \right)\nonumber\\ &&- 
    \frac{12n^2-9}{8\sqrt{3}n^2-12n}-\frac{3}{4n}\ln\left(
    \frac{2\sqrt{3}n-3}{4n^2-3}
    \right).
\end{eqnarray}

Equaling both expressions of $N_*(n)$, we obtain:
\begin{widetext}
\begin{eqnarray}\label{n_s}
    \frac{2}{1-n_s}
    +\frac{13-8n}{12n}\ln(1-n_s)
    -\frac{n-2}{12n}\ln\left(
    \frac{30M_{\rm pl}^4}{g_{\rm reh}\pi^2T_{\rm reh}^4(n)}
    \right) \cong  53.3928 
    -\frac{n+1}{3}\ln
\left(\frac{4n^2-2\sqrt{3}n}{4n^2-3}
\right)    \nonumber\\
    +\frac{n+1}{6n}\ln\left(
\frac{2\times 10^9}{9\pi^2}
\right)+ 
    \frac{12n^2-9}{8\sqrt{3}n^2-12n}+\frac{3}{4n}\ln\left(
    \frac{16\sqrt{3}n^2-24n}{12n^2-9}
    \right).    
    \end{eqnarray}
\end{widetext}
Therefore, for any model (with different values of $n$), we have established the relationship between the reheating temperature and the spectral index of scalar perturbations.
Specifically, for a viable range of reheating temperature values,
we can determine the corresponding range of the spectral index, which must be compared, at the $2\sigma$ CL, with Planck's data \cite{Planck:2018jri}.

Additionally, since the function
$F(n_s)\equiv\frac{2}{1-n_s}
    +\frac{13-8n}{12n}\ln(1-n_s)$
is increasing, for a given fixed value of $n$, one can conclude that the higher the reheating temperature, the lower the spectral index.
   
Taking into account the bound
of the reheating temperature
$
1\mbox{ MeV}\leq T_{\rm reh}(n)\leq 10^9 \mbox{ GeV}\Longleftrightarrow
5\times 10^{-22} M_{\rm pl}\leq T_{\rm reh}(n)\leq
5\times 10^{-10} M_{\rm pl}$
-- which means that it
remains within a range consistent with the Big Bang Nucleosynthesis, which occurs around the scale of $1$ MeV,  and with an upper bound around $10^9$ GeV 
in order to
avoid issues related to the gravitino problem~\cite{Ellis:1982yb,Khlopov:1984pf} --
we obtain the following range of the viable values for the spectral index depicted in \ref{table1}.

\begin{table}
\centering
\renewcommand{\arraystretch}{1.5}
\begin{tabular}{l @{\hspace{2cm}} c}
\toprule
\hline 
\textbf{$n$} & \textbf{$n_s$} \\
\hline \hline
$ 3  $ & $0.9667\leq n_s\leq 0.9683$  \\

$ 4   $ & $0.9670\leq n_s\leq 0.9693$ \\

$ 5   $ & $  0.9671\leq n_s\leq 0.9699
$ \\ 

$ 6 $ & $ 0.9672\leq n_s\leq 0.9702
$ \\

$ \infty   $ & $ 0.9677\leq n_s\leq 0.9719$ \\ 
\hline
\end{tabular}
\caption{Range of viable values of the spectral index of scalar perturbations.}
\label{table1}
\end{table}

Note that the Planck's data 
\cite{Planck:2018jri} provides 
$n_s=0.9649\pm 0.0042$ at $1\sigma$ CL, which means that the reheating bounds, constrain the Planck result.

\section{ Reheating formulas}
\label{sec-IV-application}

{
To understand the gravitational reheating mechanism, we 
 study in detail the decay process,  using the dynamics described by the Boltzmann equations:  

 \begin{eqnarray}\label{Boltzmann}
\left\{\begin{array}{ccc}
       \frac{d \langle \rho(t)\rangle}{dt}+3H\langle \rho(t)\rangle=-\Gamma \langle \rho(t)\rangle,\\
       \bigskip \\
       \frac{d \langle \rho_r(t)\rangle}{dt}+4H\langle \rho_r(t)\rangle=\Gamma \langle \rho(t) \rangle,
    \end{array}\right.
\end{eqnarray}
where
$\langle\rho(t)\rangle$
denotes the energy density of the produced particles,  
$\langle \rho_r(t)\rangle$ being the energy density of the decay products, i.e., the energy density of radiation,
and the constant decay rate
 $\Gamma$
 represents the decay of heavy particles into light fermions. Specifically, 
   $\Gamma \sim \frac{h^2 m_{\chi}}{8\pi}$ in which 
$h$ is a dimensionless constant
\cite{Felder:1999pv}.

We choose the following as the solution for the energy density of the massive particles:
\begin{eqnarray}
    \langle\rho(t)\rangle=
    \langle\rho_{\rm END}\rangle\left(
  \frac{a_{\rm END}}{a(t)}  \right)^3
  e^{-\Gamma(t-t_{\rm END})},  \quad t\geq t_{\rm END},
\end{eqnarray}
which assumes that the decay begins at the end of inflation.

Inserting this solution into the second equation of (\ref{Boltzmann}) and imposing once again that the decay starts
at the end of inflation,
we obtain:
\begin{align}\label{energydensityradiation0}    \langle\rho_{\rm r}(t)\rangle=\langle\rho_{\rm END}\rangle\left(
  \frac{a_{\rm END}}{a(t)}  \right)^4
  \int_{t_{\rm END}}^t \frac{a(s)}{a_{\rm END}}\Gamma
  e^{-\Gamma(s-t_{\rm END})}ds.
\end{align}

We define the time at which decay ends, denoted as $t_{\rm dec}$,  as the time when $\Gamma(t_{\rm dec}-t_{\rm END})\sim 1$, which gives
$t_{\rm dec}\sim t_{\rm END}+\frac{1}{\Gamma}$, and we want
to examine the evolution of the decay products when $t \gg t_{\rm dec}$. 
Assuming that the background dominates, we will have
$a(s)\cong a_{\rm END}\left(\frac{s}{t_{\rm END}} \right)^{\frac{n+1}{3n}}$, and taking into account that 
\begin{align}
&& \int_{t_{\rm END}}^{\infty}
\left(s/t_{\rm END} \right)^{\frac{n+1}{3n}}\Gamma
e^{-\Gamma(s-t_{\rm END})}ds \nonumber\\  &&\cong 
\left( \frac{H_{\rm END}}{\Gamma}\right)^{\frac{n+1}{3n}}\Gamma_{\rm Euler}
\left( \frac{4n+1}{3n} \right)\cong \left( \frac{H_{\rm END}}{\Gamma}\right)^{\frac{n+1}{3n}}
,
\end{align}
where we have denoted by $\Gamma_{\rm Euler}$ the Euler's Gamma function in order to avoid confusion with the decay rate, we  conclude that
 after the decay, i.e., for $t>t_{\rm dec}$,  we can safely make the following approximation:
\begin{eqnarray}\label{energydensityradiation}
\langle\rho_{\rm r}(t)\rangle\cong\langle\rho_{\rm END}\rangle\left( \frac{H_{\rm END}}{\Gamma}\right)^{\frac{n+1}{3n}}
\left(
  \frac{a_{\rm END}}{a(t)}  \right)^4.
\end{eqnarray}

\subsection{Reheating temperature}

We start with the so-called {\it heating efficiency} \cite{Rubio:2017gty} defined as: 
\begin{eqnarray}\label{relation}
    \Theta\equiv 
\frac{\langle\rho_{\rm END}\rangle}{3H_{\rm  \rm END}^2M_{\rm pl}^2}.
\end{eqnarray}
Using 
that
$
H_{\rm END}\cong
5\left(1- \frac{2\sqrt{3}n-3}{4n^2-3}    \right)^{n} 10^{-6}M_{\rm pl}$, which for all values of $n$ is close to $2\times 10^{-6}M_{\rm pl}$, and 
the formula of the energy density of the produced particles at the end of inflation, i.e,
$\langle \rho_{\rm END}\rangle\cong\frac{1}{4\pi^3}m_{\chi}^2H^2_{\rm END}\sqrt{\frac{m_{\chi}}{\sqrt{2}H_{\rm END}}}
$,
for $m_{\chi}\ll H_{\rm END}$,  leads to:
\begin{eqnarray}\label{theta}
&& \Theta=\frac{1}{12\pi^3}
    \left(\frac{m_{\chi}}{M_{\rm pl}}\right)^2\sqrt{\frac{m_{\chi}}{\sqrt{2}H_{\rm END}}}
    \nonumber\\ && =\frac{1}{12\sqrt{5\sqrt{2}}\pi^3}\times 10^3\left(\frac{m_{\chi}}{M_{\rm pl}}\right)^{5/2}
    \left( \frac{4n^2-3}    {4n^2-2\sqrt{3}n}\right)^{n/2}
    \nonumber\\
&& \cong\left(\frac{m_{\chi}}{M_{\rm pl}}\right)^{5/2}
   \left( \frac{4n^2-3}    {4n^2-2\sqrt{3}n}\right)^{n/2} 
 \cong 
 {1.5}
 \left(\frac{m_{\chi}}{M_{\rm pl}}\right)^{5/2}.   
 \end{eqnarray}
On the other hand, 
since  the effective EoS parameter is $w_{\rm eff}=\frac{n-1}{n+1}$,
the evolution of the background is given by
\begin{align}
    \rho_{\rm B}(t)=
     3H_{\rm  END}^2M_{\rm pl}^2\left( \frac{a_{\rm END}}{a(t)}\right)^{\frac{6n}{n+1}},
    \end{align}
and  the 
 universe becomes reheated when
$\rho_{\rm B}\sim \langle\rho_{\rm r} \rangle$, we get:
\begin{eqnarray}\label{relation}
    \Theta
= \left( \frac{\Gamma}{H_{\rm END}}\right)^{\frac{n+1}{3n}}    \left(\frac{a_{\rm END}}{a_{\rm reh}} \right)^
    {\frac{2(n-2)}{n+1}},
\end{eqnarray}
and thus, the energy density of the decay products at the reheating time is:
\begin{eqnarray}
\label{energydensity}
   \langle\rho_{\rm r, \rm reh}\rangle\cong
   3H_{ \rm END}^2M_{\rm pl}^2 
   \left(\frac{H_{\rm END}}{\Gamma} \right)^{\frac{n+1}{n-2}}
   \Theta^{\frac{3n}{n-2}}.
\end{eqnarray}

Therefore, from the
 Stefan-Boltzmann law
 $T_{\rm reh}(n)=\left( \frac{30}{\pi^2 g_{\rm reh}}\right)^{1/4}\langle\rho_{\rm r, \rm reh}\rangle^{1/4}$,
 where $g_{\rm reh}=106.75$ 
 represents the effective number of degrees of freedom in SM, 
 we obtain the following reheating temperature:
\begin{eqnarray}
\label{reheating_temperature}
&&T_{\rm reh}(n)
\cong 5\times 10^{-1}
\left(\frac{H_{\rm END}}{\Gamma} \right)^{\frac{n+1}{4(n-2)}}\Theta^{\frac{3n}{4(n-2)}}\sqrt{H_{\rm END}M_{\rm pl}}
\nonumber\\   &&\cong 
{10^{-3}}
\left(\frac{H_{\rm END}}{\Gamma} \right)^{\frac{n+1}{4(n-2)}}
\left(\frac{m_{\chi}}{M_{\rm pl}} \right)^{\frac{15n}{8(n-2)}}
M_{\rm pl}. 
\end{eqnarray}

We also need to determine the range of values for the decay rate 
$\Gamma$. Since the decay occurs well after the end of inflation,  we  have $\Gamma\ll H_{\rm END}$. On the other hand, we have assumed that by the end of the decay, the energy density of the background dominates, i.e.,
$\langle \rho_{\rm r,\rm  dec}\rangle\ll \rho_{\rm B, \rm dec}\cong 3M_{\rm pl}^2\Gamma^2$. To get the bound, we use that the relation 
$\rho_{\rm B, \rm dec}\cong 3 M_{\rm pl}^2\Gamma^2$, leads to:
\begin{eqnarray}
    \left(\frac{a_{\rm END}}{a_{\rm dec}} \right)^4\cong 
    \left( \frac{\Gamma}{H_{\rm END}}\right)^{\frac{4(n+1)}{3n}},
\end{eqnarray}
and thus, inserting it into the relation 
$\langle \rho_{\rm r,\rm  dec}\rangle\ll  3M_{\rm pl}^2\Gamma^2$, we conclude that:
\begin{eqnarray}\label{Constraint}
    \Theta^{\frac{n}{n-1}}    \ll \frac{\Gamma}{H_{\rm END}}\ll 
    1.
\end{eqnarray}

In addition, the combination 
of (\ref{Constraint})
 with the bound of the reheating temperature
$5\times 10^{-22} M_{\rm pl}\leq T_{\rm reh}(n)\leq
5\times 10^{-10} M_{\rm pl}$ 
leads to four different cases.
However,  we have checked that the only viable case {for $n\geq 3$} is when 
\begin{widetext}
\begin{eqnarray}\label{constraint0}
    10^{\frac{36(n-2)}{n+1}}\left(
    \frac{H_{\rm END}}{M_{\rm pl}}
    \right)^{ \frac{2(n-2)}{n+1} }\Theta^{\frac{3n}{n+1}}
    \leq 
    \Theta^{\frac{n}{n-1}}\ll
    \frac{\Gamma}{H_{\rm END}}\leq 
     10^{\frac{84(n-2)}{n+1}}\left(
    \frac{H_{\rm END}}{M_{\rm pl}}
    \right)^{ \frac{2(n-2)}{n+1} }\Theta^{\frac{3n}{n+1}}\ll 1,
\end{eqnarray}
\end{widetext}
{
provided that:
\begin{align}
     10^{-\frac{42(n-1)}{n}}\left(
    \frac{M_{\rm pl}}
    {H_{\rm END}}
    \right)^{ \frac{n-1}{n} }
    \ll
    \Theta\ll
     10^{-\frac{28(n-2)}{n}}\left(
    \frac{M_{\rm pl}}
    {H_{\rm END}}
    \right)^{ \frac{2(n-2)}{3n} }.
\end{align}
Inserting the 
 values of 
$H_{\rm END}\cong 2\times 10^{-6}M_{\rm pl}$ and $\Theta\cong
{1.5}
\left(m_{\chi}/M_{\rm pl}\right)^{5/2}$, one gets that the reheating temperature is given by (\ref{reheating_temperature})
provided that:
\begin{widetext}
\begin{eqnarray}\label{constraint}
    10^{-\frac{14(n-1)}{n}}    \ll
    \frac{m_{\chi}}{M_{\rm pl}}\ll  {
  \mbox{min} \left( 10^{-\frac{48(n-2)}{5n}}; 2\times 10^{-6} \right)}
\quad\mbox{ and }\quad
\left(
    \frac{m_{\chi}}{M_{\rm pl}}\right)^{\frac{5n}{2(n-1)}}\ll
    \frac{\Gamma}{H_{\rm END}}\lesssim { 20}\times
     10^{\frac{72(n-2)}{n+1}}
     \left(\frac{m_{\chi}}{M_{\rm pl}}\right)^{\frac{15n}{2(n+1)}}     .
\end{eqnarray}
\end{widetext}
\begin{remark}
From the previous calculations, one can show that the reheating temperature (\ref{reheating_temperature}) is bounded by: 
\begin{widetext}
\begin{eqnarray}
\label{bound1}
    5\times 10^{-22}M_{\rm pl}\leq T_{\rm reh}(n)\ll 
    {10^{-3}}
    \left( \frac{m_{\chi}}{M_{\rm pl}}\right)^{\frac{5n}{4(n-1)}}M_{\rm pl}
    \ll { 10^{-15}}
    M_{\rm pl}.
\end{eqnarray}
\end{widetext}
\end{remark}
\begin{table*}
\renewcommand{\arraystretch}{1.1}
\resizebox{0.9\textwidth}{!}
{\begin{tabular}{l @{\hspace{1cm}} c @{\hspace{1cm}} c @{\hspace{1cm}} c}
\toprule
\hline 
$n$ & $m_{\chi}/M_{\rm pl}$ & $\frac{\Gamma}{H_{\rm END}}\left(\frac{m_{\chi}}{M_{\rm pl}}\right)
$& 
$\frac{T_{\rm reh}}{M_{\rm pl}}
\left(\frac{m_{\chi}}{M_{\rm pl}}, 
\frac{\Gamma}{H_{\rm END}}\right)$ \\
\hline \hline

$ 3  $ &  $  5\times 10^{-10}\ll {m_{\chi}}/{M_{\rm pl}}\ll {2\times 10^{-6} }
\quad   $ & $  
  10^{-30}\ll
\frac{\Gamma}{H_{\rm END}}(10^{-8})
\lesssim 2\times 10^{-26} 
  $ & $   
\frac{T_{\rm reh}}{M_{\rm pl}}(10^{-8}, 10^{-27})\cong 10^{-21}$ \\ 

$ 4   $  & $  3\times 10^{-11}\ll {m_{\chi}}/{M_{\rm pl}}\ll {2\times 10^{-6}}
  $ & $  
10^{-30}\ll
\frac{\Gamma}{H_{\rm END}}(10^{-9})
\lesssim 10^{-24}$ & $  
\frac{T_{\rm reh}}{M_{\rm pl}}(10^{-9},10^{-29})\cong  2\times 10^{-19}$ \\

$ 5   $ & $  6\times 10^{-12}\ll {m_{\chi}}/{M_{\rm pl}}\ll  { 2\times 10^{-6}}$ & $  
7\times 10^{-29}\ll
\frac{\Gamma}{H_{\rm END}}(10^{-9})
\lesssim  10^{-19}$ & $  
\frac{T_{\rm reh}}{M_{\rm pl}}(10^{-9},10^{-27})\cong 2\times 10^{-18}$ \\ 

$ 6   $ & $  2\times 10^{-12}\ll {m_{\chi}}/{M_{\rm pl}}\ll {4\times 10^{-7}}$ & $  
 10^{-30}\ll
\frac{\Gamma}{H_{\rm END}}(10^{-10})
\lesssim  10^{-22}$ & $  
\frac{T_{\rm reh}}{M_{\rm pl}}(10^{-10},10^{-28})\cong  10^{-19}$ \\ 

$ \infty   $ & $  10^{-14}\ll {m_{\chi}}/{M_{\rm pl}}\ll  { 3\times 10^{-10}}$ & $  
 10^{-30}\ll
\frac{\Gamma}{H_{\rm END}}(10^{-12})
\lesssim 2\times 10^{-17}$ & $  
\frac{T_{\rm reh}}{M_{\rm pl}}(10^{-12},10^{-28})\cong  3\times 10^{-19}$  \\ 
\hline 
\end{tabular}}
\caption{Range of viable masses, range of the decay rate for a fixed mass and the reheating temperature for a fixed mass and decay rate.}
\label{table2}
\end{table*}
In Table~\ref{table2}, 
we calculate, for some values of the parameter $n$, the corresponding range of masses and
fixing the mass we provide bounds for the decay rate. In addition, fixing these two parameters we calculate the reheating temperature, showing that  for many values of the viable masses, it belongs the the MeV regime. This means that very constrained bounds -- assuming  reheating  via the decay of the inflaton's field -- provided by the gravitino problem
\cite{Khlopov:1984pf,Kawasaki:2004qu,Kawasaki:2006hm,Kawasaki:2017bqm} were over-passed. In addition,   baryogenenis could work as discussed in~\cite{Davidson:2000dw}.

To end this discussion, we deal with  inflation followed by kination (a regime where all the energy density of the inflaton field is kinetic) \cite{Peebles:1998qn,Giovannini:1998bp}, that is, with the case $n=\infty$. The reheating temperature takes the form
\begin{eqnarray}
&& T_{\rm reh}(\infty)
\cong 5\times 10^{-1}
\left(\frac{H_{\rm END}}{\Gamma} \right)^{\frac{1}{4}}\Theta^{\frac{3}{4}}\sqrt{H_{\rm END}M_{\rm pl}}
\nonumber\\
&&\cong 
{9\times 10^{-4}}
\left(\frac{H_{\rm END}}{\Gamma} \right)^{\frac{1}{4}}
\left(\frac{m_{\chi}}{M_{\rm pl}
} \right)^{\frac{15}{8}}M_{\rm pl},\end{eqnarray}
with the constraints:

\begin{eqnarray}
\left\{\begin{array}{ccc}
        10^{-14}
    \ll
    \frac{m_{\chi}}{M_{\rm pl}}\ll  3\times 10^{-10}, \\
    \bigskip \\
       \left(
    \frac{m_{\chi}}{M_{\rm pl}}\right)^{\frac{5}{2
    }}\ll
    \frac{\Gamma}{H_{\rm END}}\lesssim {2\times 10^{73}}
\left(\frac{m_{\chi}}{M_{\rm pl}}\right)^{\frac{15}{2}},
    \end{array}\right.
\end{eqnarray}
which leads to the bound:

\begin{eqnarray}
\left\{\begin{array}{ccc}
5\times 10^{-22}M_{\rm pl}\lesssim T_{\rm reh}(\infty)\ll  10^{-3}
\left(\frac{m_{\chi}}{M_{\rm pl}}\right)^{\frac{5}{4}}, \\
    \bigskip \\
{\rm with}\; \; \;  10^{-14}
    \ll
    \frac{m_{\chi}}{M_{\rm pl}}\ll  3\times10^{-10}
    \end{array}\right.
\end{eqnarray}
In fact, the greatest value of the reheating temperature is obtained when 
\begin{eqnarray}
\frac{m_{\chi}}{M_{\rm pl}}\sim  10^{-11}, \qquad\mbox{and}\qquad
    \frac{\Gamma}{H_{\rm END}}\sim 
    10\left(
    \frac{m_{\chi}}{M_{\rm pl}}\right)^{\frac{5}{2
    }},\end{eqnarray}
leading to $T_{\rm reh}(\infty)\sim 10^{-17}M_{\rm pl}\cong 2\times 10$ GeV, which is a very low temperature, and thus, 
over-passing  the gravitino problem.

}

\subsection{Maximum reheating temperature}\label{maximum}
}

\begin{figure}
    \centering
    \includegraphics[width=0.5\textwidth]{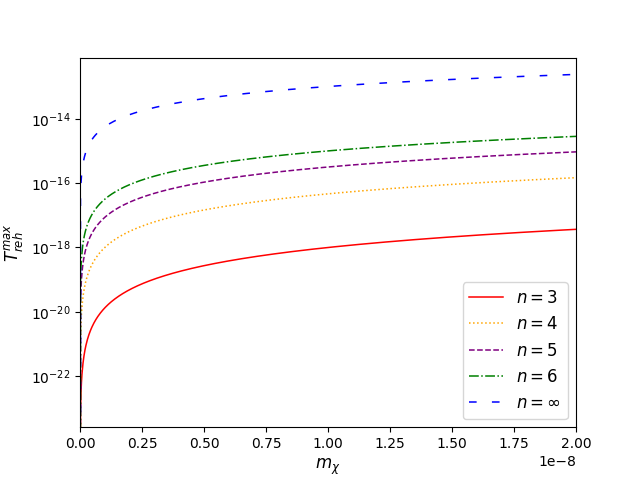}
    \caption{We show the nature of $T_{\rm max}^{\rm reh}$ influenced by the mass of the produced particles, $m_{\chi}$, for different values of $n$, considering the units where $M_{\rm pl} =1$. }
    \label{fig:Treh-vs-mx}
\end{figure}

As we discussed in \cite{deHaro:2024xgd}, the maximum value of the reheating temperature is obtained when the decay coincides with the end of the inflaton's domination, that is, when $\langle \rho_{\rm r,\rm reh}\rangle
\sim 3\Gamma^2M_{\rm pl}^2$. 
{ 
This is achieved when there is no delay in the decay (the delayed decay was studied at the beginning of this section), meaning that the decay is nearly instantaneous. As demonstrated in \cite{deHaro:2024xgd}, the decay coincides with the end of the inflaton's domination when
\begin{eqnarray}
    \frac{\Gamma}{H_{\rm END}}\sim \Theta^{\frac{n}{n-1}},
\end{eqnarray}
which leads to the maximum reheating temperature (Eq. (29) of \cite{deHaro:2024xgd}):
\begin{eqnarray}\label{Tem_max}
    T_{\rm reh}^{\rm max}(n)
\cong 5\times 10^{-1}
\Theta^{\frac{n}{2(n-1)}}\sqrt{H_{\rm END}M_{\rm pl}}.\end{eqnarray}
} Therefore, 
using the formula of the energy density of the produced particles at the end of inflation, i.e,
$\langle \rho_{\rm END}\rangle\cong\frac{1}{4\pi^3}m_{\chi}^2H^2_{\rm END}\sqrt{\frac{m_{\chi}}{\sqrt{2}H_{\rm END}}}
$,
for $m_{\chi}\ll H_{\rm END}$, we have:
\begin{eqnarray}
    T_{\rm reh}^{\rm max}(n)\cong 
    10^{-3}
    \left(\frac{m_{\chi}}{M_{\rm pl}}\right)^{\frac{5n}{4(n-1)}}M_{\rm pl},
\end{eqnarray}
and the bound
$5\times 10^{-22} M_{\rm pl}\leq T_{\rm reh}^{\rm max}(n)\leq
5\times 10^{-10} M_{\rm pl}$
leads to the constraint:
\begin{eqnarray}
    2\times 10^{-\frac{15(n-1)}{n}}
    \lesssim    
\frac{m_{\chi}}{M_{\rm pl}}
     \ll 2\times 10^{-6}
     ,
    \end{eqnarray}
where we have taken into account that
$  
10^{-\frac{28(n-1)}{5n}}
   \geq 3\times 10^{-6} M_{\rm pl}\sim \frac{H_{\rm END}}{M_{\rm pl}} $, for $n>2$. Then,  the maximum reheating temperature and the 
   constraint become:
\begin{eqnarray}\label{T_max}
\left\{\begin{array}{ccc}
    T_{\rm reh}^{\rm max}(n)\cong 
    10^{-3}
    \left(\frac{m_{\chi}}{M_{\rm pl}}\right)^{\frac{5n}{4(n-1)}}M_{\rm pl}, \\
    \bigskip \\
      {\rm with}\;\;2\times 10^{-\frac{15(n-1)}{n}}
    \lesssim \frac{m_{\chi}}{M_{\rm pl}}
    \ll 2\times 10^{-6}.
    \end{array}\right.
\end{eqnarray}
In Table~\ref{table3}, 
taking different values of $n$, 
we show the  viable values of the mass and the upper bound of the maximum reheating temperature.
And in Fig. \ref{fig:Treh-vs-mx},  we present the graphical nature of $T_{\rm reh}^{\rm max}$ {\it versus} $m_{\chi}$ for different values of $n$.

\begin{table}
\begin{center}
\renewcommand{\arraystretch}{1.5}
\resizebox{0.5\textwidth}{!}
{
\begin{tabular}{c
@{\hspace{0.75cm}}c
@{\hspace{0.5cm}} c}
\hline 
$n$ & $m_{\chi}/M_{\rm pl}$ & 
${T_{\rm reh}^{\rm max}}/{M_{\rm pl}}
$  \\ 
\hline 
\hline
$ 3  $ &  $  2\times 10^{-10}\leq {m_{\chi}}/{M_{\rm pl}}\ll 2\times 10^{-6} 
\quad   $ &  $   
{T_{\rm reh}^{\rm max}}/{M_{\rm pl}}\ll 2\times 10^{-14}$ \\ 

$ 4   $  & $   10^{-11}\leq {m_{\chi}}/{M_{\rm pl}}\ll 2\times 10^{-6}
  $ &  $  
{T_{\rm reh}^{\rm max}}/{M_{\rm pl}}\ll 3\times 10^{-13} $ \\ 

$ 5   $ & $  2\times 10^{-12}\leq {m_{\chi}}/{M_{\rm pl}}\ll 2\times 10^{-6}$ &  $  
{T_{\rm reh}^{\rm max}}/{M_{\rm pl}} \ll 10^{-12}$ \\ 

$ 6   $ & $  6\times 10^{-13}\leq {m_{\chi}}/{M_{\rm pl}}\ll 2\times 10^{-6}$ &  $  
{T_{\rm reh}^{\rm max}}/{M_{\rm pl}}\ll 3\times 10^{-12}$ \\ 

$ \infty   $ & $  2\times 10^{-15}\leq {m_{\chi}}/{M_{\rm pl}}\ll 2\times 10^{-6}$ &  $  
{T_{\rm reh}^{\rm max}}/{M_{\rm pl}}\ll 7\times 10^{-11} $  
\\ 
\hline 
\end{tabular} }
\end{center}
\caption{Range of viable masses and upper bound of the maximum reheating temperature. 
}
\label{table3}
\end{table}

An important final remark is in order:
Since  the maximum reheating temperature 
only depends on the mass of produced particles, one has a direct relationship between  that  mass and the spectral index. And recalling that for the range of viable masses (\ref{T_max}), the maximum reheating temperature
is lower than $10^9$ GeV, we can better constraint the spectral index,
as we show in Table~\ref{table4}. 

\begin{table}
\begin{center}
\renewcommand{\arraystretch}{1.5}
\resizebox{0.5\textwidth}{!}
{
\begin{tabular}{c@{\hspace{1cm}} c
@{\hspace{1cm}} c}
\hline 
$n$ & $n_s$ & 
${T_{\rm reh}^{\rm max}}/{M_{\rm pl}}
$
\\ 
\hline
\hline 

$ 3  $ &  $  0.9677 \leq n_s\leq 0.9683   $ & $   
{T_{\rm reh}^{\rm max}}/{M_{\rm pl}}\leq 2\times 10^{-15}$ \\

$ 4   $  & $   0.9678\leq n_s\leq 0.9693
  $ &  $  
{T_{\rm reh}^{\rm max}}/{M_{\rm pl}}\leq 3\times  10^{-14}$ \\ 

$ 5   $ & $  0.9679\leq n_s\leq 0.9699
$ & $  
{T_{\rm reh}^{\rm max}}/{M_{\rm pl}}\leq 10^{-13} $ \\

$ 6   $ & $ 0.9681\leq n_s \leq 0.9702
$ &  $  
{T_{\rm reh}^{\rm max}}/{M_{\rm pl}}\leq  3\times 10^{-13}$ \\ 

$ \infty   $ & $ 0.9684\leq n_s\leq 0.9719$  & $  
{T_{\rm reh}^{\rm max}}/{M_{\rm pl}}\leq 7\times 10^{-12} $  \\
\hline 
\end{tabular} }
\end{center}
\caption{ Range of viable values of the spectral index when one considers the maximum reheating temperature. 
}
\label{table4}
\end{table}

{
\subsection{Decay after inflaton's domination}

Assuming that the decay is well after the background domination, this implies that $t_{\rm end}\Gamma\ll 1$.
In this situation, during inflaton's domination, the evolution of the energy density of the produced particles is:
\begin{eqnarray}
    \langle \rho(t)\rangle\cong \langle \rho_{\rm END}\rangle
    \left( \frac{a_{\rm END}}{a(t)}\right)^3,
    \end{eqnarray}
and thus, from formula (32) of 
\cite{deHaro:2024xgd}
we have $H_{\rm end}=
\sqrt{2}\Theta^{\frac{n}{n-1}}H_{\rm END}$. Therefore, since $t_{\rm end}\sim 1/H_{\rm end}$, we have the following constraint:
\begin{eqnarray}\label{boundafter}
\frac{\Gamma}{H_{\rm END}}\ll \sqrt{2}\Theta^{\frac{n}{n-1}}.
\end{eqnarray}
On the other hand, when the decay is nearly instantaneous, 
the reheating temperature is given by
\begin{eqnarray}
 T_{\rm reh}(n)=\left(\frac{30}{\pi^2 g_{\rm reh}}\right)^{1/4} 
 \langle\rho_{r,\rm dec}\rangle^{1/4}.
\end{eqnarray}
This expression can be improved taking into account 
that
$3\Gamma^2M_{\rm pl}^2\cong \langle \rho_{r,\rm dec}\rangle$, which leads to:
\begin{eqnarray}
T_{\rm reh}(n)\cong 5\times 10^{-1}\sqrt{\Gamma M_{\rm pl}},
\end{eqnarray}
and from the constraint
(\ref{boundafter}), we find the upper bound:
\begin{eqnarray}
T_{\rm reh}(n)\ll 6\times 10^{-1}
\Theta^{\frac{n}{2(n-1)}}
\sqrt{H_{\rm END} M_{\rm pl}}.
\end{eqnarray}

Finally, using the values of 
$H_{\rm END}\cong 2\times 10^{-6}M_{\rm pl}$ and $\Theta\cong 1.5 \left(m_{\chi}/M_{\rm pl}\right)^{5/2}$, we obtain:
\begin{eqnarray}
T_{\rm reh}(n)\ll  10^{-3}
 \left( \frac{m_{\chi}}{M_{\rm pl}}\right)^{\frac{5n}{4(n-1)}} M_{\rm pl},
\end{eqnarray}
which 
coincides with the upper bound in  (\ref{bound1}).
In contrast, when the decay is delayed, the situation is more involved. Effectively, first of all, from the formula (\ref{energydensityradiation0}),  we have to calculate the evolution of the energy density of radiation. We split it, for $t>t_{\rm end}$,  as follows:
\begin{align}
\label{energydensityradiation2}
  \langle \rho_r(t)\rangle=
    \langle\rho_{\rm END}\rangle\left(\frac{a_{\rm END}}{a(t)} \right)^4\left[ \int_{t_{\rm END}}^{t_{\rm end}}
    \frac{a(s)}{a_{\rm END}}\Gamma e^{-\Gamma(s-t_{\rm END})}ds\right. \nonumber\\ \left.+
    \int_{t_{\rm end}}^{t}
    \frac{a(s)}{a_{\rm END}}\Gamma e^{-\Gamma(s-t_{\rm END})}ds    \right].
\end{align}
In the first integral we
use the approximation $a(s)\cong a_{\rm END}\left( \frac{s}{t_{\rm END}}\right)^{\frac{n+1}{3n}}$ leading to a value of the order 
$\frac{\Gamma}{H_{\rm end}}\left( \frac{H_{\rm END}}{H_{\rm end}}\right)^{\frac{n+1}{3n}}=\frac{\Gamma}{H_{\rm end}}\Theta^{-\frac{n+1}{3(n-1)}}$.
On the other hand, to evaluate the second integral, we assume that after $t_{\rm end}$ the leading terms of the total energy density is $\langle \rho(t)\rangle$. Then, 
\begin{eqnarray}
    \frac{\dot{a}(t)}{a(t)}\cong
   \sqrt{\Theta} H_{\rm END}\left(
  \frac{a_{\rm END}}{a(t)}  \right)^{3/2}
  e^{-\frac{\Gamma}{2}t},\end{eqnarray}
which leads to 
\begin{align}
    \frac{a(t)}{a_{\rm END}}\cong
\left(\frac{3H_{\rm END}\sqrt{\Theta}}{\Gamma} \right)^{2/3}\left[ 1-e^{-\frac{\Gamma}{2}t}
+\frac{\Gamma}{3H_{\rm END}}
\Theta^{-\frac{n}{n-1}}
    \right]^{2/3},
\end{align}
and this under the
condition $\frac{\Gamma}{H_{\rm END}}\ll \Theta^{\frac{n}{n-1}}$
can be approximated by
\begin{align}\label{scale_factor}
    \frac{a(t)}{a_{\rm END}}\cong
\left(\frac{3H_{\rm END}\sqrt{\Theta}}{\Gamma} \right)^{2/3}\left[ 1-e^{-\frac{\Gamma}{2}t}
    \right]^{2/3}.
\end{align}
Therefore, one has 
\begin{align}
& \int_{t_{\rm end}}^t \frac{a(s)}{a_{\rm END}}\Gamma_{\varphi}
  e^{-\Gamma(s-t_{\rm END})}ds
 \nonumber\\ 
& \cong  \left( \frac{3H_{\rm END}\sqrt{\Theta}}{\Gamma}\right)^{2/3}
\int_0^{\Gamma t}
  \left(1-e^{-x/2} \right)^{2/3}
  e^{-x}dx,
\end{align}
where we have used that $t_{\rm END}\Gamma\ll
t_{\rm end}\Gamma\ll
1$.
This integral can be analytically solved, obtaining:
\begin{align}
& \int_{t_{\rm end}}^t \frac{a(s)}{a_{\rm END}}\Gamma_{\varphi}
  e^{-\Gamma(s-t_{\rm END})}ds\nonumber\\
  & \cong \frac{3}{20}
  \left(\frac{3H_{\rm END}\sqrt{\Theta}}{\Gamma}\right)^{2/3}  \left(1-e^{-\Gamma t/2} \right)^{5/3}
    \left(3+5e^{-\Gamma t/2}\right).\end{align}
This quantity is of the order $\left(\frac{H_{\rm END}\sqrt{\Theta}}{\Gamma}\right)^{2/3}$, and has to be compared with $\frac{\Gamma}{H_{\rm end}}\Theta^{-\frac{n+1}{3(n-1)}}$ coming from the first integral in (\ref{energydensityradiation2}). The bound (\ref{boundafter}) ensures us that the dominant term comes from the second integral of (\ref{energydensityradiation2}). Therefore,
\begin{align}\label{radiation_formula}
\langle\rho_{ r}(t)\rangle\cong
\frac{3}{20}\langle\rho_{\rm  END}\rangle
\left( \frac{3H_{\rm END}\sqrt{\Theta}}{\Gamma}\right)^{2/3}
\left(1-e^{-t\Gamma/2} \right)^{5/3}\nonumber\\
  \times  \left(3+5e^{-t\Gamma/2}\right)
\left(
  \frac{a_{\rm END}}{a(t)}  \right)^4,
\end{align}
which in order to obtain the reheating temperature has to be equaled to $\langle \rho(t)\rangle$ at time $t_{\rm reh}$.

Introducing the notation,
$z_{\rm reh}\equiv e^{-t_{\rm reh}\Gamma/2}$, we find:
\begin{eqnarray}
    35z_{\rm reh}^2-6z_{\rm reh}-9=0,
\end{eqnarray}
whose solution is $z_{\rm reh}=3/5$, and leads to the following energy density
$\langle\rho_{r,\rm reh}\rangle\cong \frac{3}{4}
\Gamma^2
M_{\rm pl}^2
$. Therefore, the reheating temperature is:
\begin{align}
    T_{\rm reh}(n)=
    \left(\frac{30}{\pi^2g_{\rm reh}}\right)^{1/4}
    \langle\rho_{r,\rm reh}\rangle^{1/4} \cong 4\times 10^{-1}\sqrt{\Gamma M_{\rm pl}}.   \end{align}


\subsection{Constraints coming from the overproduction of Gravitational Waves}
\label{GW}

Since $w_{\rm eff}=\frac{n-1}{n+1}$,
for large values of $n$, the effective EoS parameter is close to $1$ as in Quintessential Inflation, and thus the problem of the overproduction
of Graviational Waves (GW)
at the end of inflation appears.

The success of the BBN imposes the constraint~\cite{Peebles:1998qn}
\begin{align}\label{GWconstraint}
    \frac{\rho_{\rm GW, \rm reh}}{\langle\rho_{r,\rm reh}\rangle}\leq 7\times 10^{-2},
\end{align}
where $\rho_{\rm GW}(t)\cong 10^{-2}H_{\rm END}^4\left(a_{\rm END}/a(t) \right)^4$  is the energy density of the GW.
To deal with the consequences of this bound on the value of the reheating temperature, we start considering a delayed decay during the inflaton's domination.
Taking into account that we are considering large values of the parameter $n$, the energy density of radiation at the reheating time is given by (\ref{energydensity}):
\begin{align}
    \langle \rho_{r,\rm reh}\rangle\cong 3H_{\rm END}^2M_{\rm pl}^2\frac{H_{\rm EMD}}{\Gamma}\Theta^3.
\end{align}
Taking into account that now
\begin{align}
    \Theta\cong \left(\frac{\Gamma}{H_{\rm END}}\right)^{1/3}\left(\frac{a_{\rm END}}{a_{\rm reh}}\right)^2, 
\end{align}
the energy density of the GW at the reheating time is
\begin{align}
    \rho_{\rm GW,\rm reh}\cong 10^{-2}H_{\rm END}^4\left(\frac{H_{\rm END}}{\Gamma}\right)^{2/3}
\Theta^2,
\end{align}
and thus, 
\begin{align}
    \frac{\rho_{\rm GW, \rm reh}}{\langle\rho_{r,\rm reh}\rangle}\cong 
    \frac{1}{3}10^{-2}
    \left(\frac{H_{\rm END}}{M_{\rm pl}} \right)^2\left(\frac{\Gamma}{H_{\rm END}}\right)^{1/3}\Theta^{-1}.
    \end{align}
Therefore, the bound (\ref{GWconstraint}) leads to:
\begin{align}\label{const}
    \frac{\Gamma}{H_{\rm END}}\leq 21^3\Theta^3\left(
    \frac{M_{\rm pl}}{H_{\rm END}}\right)^6.
\end{align}
On the other hand, the reheating temperature is given by:
\begin{align}
    T_{\rm reh}\cong 5\times 10^{-1}\left(\frac{H_{\rm END}}{\Gamma}\right)^{1/4}
\Theta^{3/4}\sqrt{H_{\rm END}M_{\rm pl}},
\end{align}
and thus, the  constraint $5\times 10^{-22}M_{\rm pl}\leq T_{\rm reh}\leq 5\times 10^{-10}M_{\rm pl}$, together with (\ref{const}) imposes:
\begin{align}
10^{36}\left(\frac{H_{\rm END}}{M_{\rm pl}}\right)^2\Theta^3
\leq \frac{\Gamma}{H_{\rm END}}\leq 21^3\left(\frac{M_{\rm pl}}{H_{\rm END}}\right)^6\Theta^3,
\end{align}
which has to be compatible with 
$\Theta\ll \frac{\Gamma}{H_{\rm END}}\ll 1$ (see
(\ref{Constraint})). 
As we have already discussed, there are four possible combinations. However, we have checked that the only viable one is the  combination of  (\ref{constraint0})  with  (\ref{const}):
\begin{align}
10^{36}\left(\frac{H_{\rm END}}{M_{\rm pl}}\right)^2\Theta^3
\leq \Theta\ll 
\frac{\Gamma}{H_{\rm END}}\leq 21^3\left(\frac{M_{\rm pl}}{H_{\rm END}}\right)^6\Theta^3
\ll 1. \end{align}
Recalling that $H_{\rm END}\cong 2\times 10^{-6} M_{\rm pl}$, we have
\begin{align}\label{constraintGW}
4\times 10^{24}\Theta^3
\leq \Theta\ll 
\frac{\Gamma}{H_{\rm END}}\leq 1.5\times 10^{38}\Theta^3
\ll 1 \end{align}
provided that: 
\begin{align}
   10^{-19}\ll
   \Theta\ll 10^{-13}.
  \end{align}
And taking into account that $\Theta\cong 1.5 \left(m_{\chi}/M_{\rm pl}\right)^{5/2}$, we find the range of viable masses:
\begin{align}
    2\times 10^{-8}\ll \frac{m_{\chi}}{M_{\rm pl}}
    \ll 2\times 10^{-6},\end{align}
which constrains the value of the viable masses to be
$m_{\chi}\cong 2\times 10^{-7}
{M_{\rm pl}}$.

Finally, we deal with the maximum reheating temperature. In this case, for large values of $n$, one has:
\begin{align}
\langle\rho_{r,\rm reh}\rangle\cong 3\Theta^2H_{\rm END}^2M_{\rm pl}^2,
\end{align}
which after inserting into (\ref{GWconstraint}), leads to:
\begin{align}
    \Theta\geq 8\times 10^{-20},
\end{align}
where we have used that when the decay is at the end of the background domination,  one has
$\Theta=\left(a_{\rm END}/a_{\rm reh}\right)^3$.
In addition, we have to take into account the bounds, coming from the success of the BBN,  for the reheating temperature
$T_{\rm reh}^{\rm max}\cong 5\times 10^{-1}\sqrt{\Theta H_{\rm END}M_{\rm pl}}$, which leads to:
\begin{align}
5\times 10^{-37}\leq \Theta\leq 5\times 10^{-13}.\end{align}
Combining both the constraints, we obtain:
\begin{align}
8\times 10^{-20}\leq \Theta\leq 5\times 10^{-13},\end{align}
with in terms of the mass of the produced particles, 
and taking into account that we consider masses below 
the Hubble rate at the end of inflation, 
becomes:
\begin{align}
    2\times 10^{-15}\leq \frac{m_{\chi}}{M_{\rm pl}}\ll 2\times 10^{-6}.
\end{align}

}

\section{Gravitational  dark matter production}
\label{sec-V}

In this section, we study the gravitational production of dark matter within the context of gravitational reheating. Specifically, we consider two quantum scalar fields, $X$ and $Y$. The $X$-field 
is responsible for the gravitational production of 
$X$-particles with mass $m_X$, which will decay into SM particles to reheat the universe. 
The $Y$-field,  on the other hand, is responsible for the production of $Y$-particles with 
 mass $m_Y$, which will account for present-day dark matter.

Assuming that the decay of the $X$-particles occurs during the inflaton domination, at the onset of radiation,  energy density of the $X$-particles is (\ref{energydensity})
\begin{eqnarray}
   \langle\rho_{X, \rm reh}\rangle\cong
   3H_{ \rm END}^2M_{\rm pl}^2 
   \left(\frac{H_{\rm END}}{\Gamma_X} \right)^{\frac{n+1}{n-2}}
   \Theta_X^{\frac{3n}{n-2}},
\end{eqnarray}  
where $\Gamma_X$ is the decay rate of the $X$-particles and $\Theta_X$ the heating efficiency corresponding to  the $X$-particles. 
At the matter-radiation equality, which we will denote by ``eq'', we will have:
\begin{eqnarray}
    \frac{a_{\rm reh}}{a_{\rm eq}}=\frac{\langle \rho_{Y,\rm reh}\rangle}{\langle \rho_{X,\rm reh}\rangle}\Longrightarrow
    \langle \rho_{Y,\rm eq}\rangle
    =\frac{\langle \rho_{Y,\rm reh}\rangle^4} {\langle \rho_{X,\rm reh}\rangle^3}.
    \end{eqnarray}

Next, we have to take into account that 
\begin{eqnarray}
&& \langle\rho_{Y, \rm reh}\rangle =
    \langle\rho_{Y, \rm END}\rangle\left(\frac{a_{\rm END}}{a_{\rm reh}} \right)^3
   \nonumber\\ &&=
   3H_{ \rm END}^2M_{\rm pl}^2 
   \left(\frac{H_{\rm END}}{\Gamma_X} \right)^{\frac{(n+1)^2}{
   2n(n-2)}}
   \Theta_X^{\frac{3(n+1)}{2(n-2)}}
   \Theta_Y,
\end{eqnarray}
where we have used the relation (\ref{relation}):
\begin{eqnarray}
    \Theta_X
= \left( \frac{\Gamma_X}{H_{\rm END}}\right)^{\frac{n+1}{3n}}    \left(\frac{a_{\rm END}}{a_{\rm reh}} \right)^
    {\frac{2(n-2)}{n+1}}.
\end{eqnarray}

Therefore:
\begin{eqnarray}
    \langle \rho_{Y,\rm eq}\rangle
    =3H_{ \rm END}^2M_{\rm pl}^2 
   \left(\frac{H_{\rm END}}{\Gamma_X} \right)^{-\frac{n+1}{
   n}}
   \Theta_X^{-3}
   \Theta_Y^4    
    .
    \end{eqnarray}

On the other hand, considering the central values  of the redshift at the matter-radiation
equality $z_{\rm eq} = 3365$, the present value of the ratio of
the matter energy density to the critical one is 
$\Omega_{\rm m,0} =0.308$, and $H_0 = 67.81$ Km/sec/Mpc\footnote{Note that the values of $\Omega_{m0}$ and $H_0$ are very much similar to Planck 2018 \cite{Planck:2018vyg}. }, one can deduce
that the present value of the matter energy density is
$\rho_{\rm m,0} = 
3H^2_0M^2_{\rm pl}\Omega_{\rm m,0} \cong 3\times 10^{-121}M^4_{\rm pl}$, and at matter
radiation equality one will have $\rho_{\rm m,\rm eq} = \rho_{\rm m,0}(1 +z_{\rm eq})^3 \cong 10^{-110}M_{\rm pl}^4$.
Since practically all the matter has a
not baryonic origin, one can conclude that $\langle\rho_{Y,\rm eq} \rangle\cong \rho_{\rm m,\rm eq}$, and thus, we have the relation between the decay rate of the $X$-particles and the heating efficiencies. 
Then we get, 
\begin{eqnarray}
    \frac{\Gamma_X}{H_{\rm END}}
    \cong 
\left(\frac{\Theta_X^3}{12\Theta_Y^4}
    \right)^{\frac{n}{n+1}}10^{-\frac{98n}{n+1}},
\end{eqnarray}
where we have used that $H_{\rm END}\cong 2\times 10^{-6} M_{\rm pl}$, and 
from 
 the constraint 
(\ref{constraint}), we get:
{
\begin{align}
    10^{99}\left(\frac{m_X}{M_{\rm pl}}\right)^{\frac{5(n+1)}{2(n-1)}}\ll \frac{\Theta_X^3}{
    \Theta_Y^4    } 
    \lesssim 5
    \times 10^{\frac{2(85n-72)}{n}}
    \left(\frac{m_X}{M_{\rm pl}}\right)^{\frac{15}{2}}.    \end{align}

Next, from (\ref{theta}), we 
have $\Theta_{A}\cong
{1.5}
\left(\frac{m_A}{M_{\rm pl}}\right)^{5/2}$, with $A=X,Y$, which leads to the following relation between the masses:

\begin{eqnarray}
   8\times 10^{-18}\times 10^{\frac{72}{5n}}
\lesssim
    \frac{m_Y}{M_{\rm pl}}
    \ll 10^{-10} 
\left(\frac{m_X}{M_{\rm pl}}
    \right)^{\frac{n-2}{2(n-1)}}.    
    \end{eqnarray}

{
It is important to recognize that this constraint holds for moderate values of $n$. However, for large values of $n$ we have to take into account the bounds coming from the overproduction of GW. In this case, the constraint (\ref{constraintGW}) leads to 
\begin{align}
    6\times 10^{-35}\leq \Theta_Y\ll 2\times 10^{-25}\sqrt{\Theta_X},
\end{align}
that is:
\begin{align}
    2\times 10^{-14}\leq\frac{m_Y}{M_{\rm pl}}\ll 10^{-10}\sqrt{\frac{m_X}{M_{\rm pl}}}\cong 4\times 10^{-14},
\end{align}
which is a contradiction. That means, the gravitational production of dark matter and the delayed decay of the particles responsible of the reheating is incompatible for large values of $n$.}

In Table~\ref{table5} we present the range of viable masses for different values of $n$.

\begin{table}[t]
\begin{center}
\renewcommand{\arraystretch}{1.5}
\resizebox{0.5\textwidth}{!}{
\begin{tabular}{c
@{\hspace{0.5cm}}
c
@{\hspace{0.5cm}}c}
\hline 
$n$ & $m_{X}/M_{\rm pl}$ & 
$m_{Y}/M_{\rm pl}$
\\ 
\hline
\hline 

$ 3  $ &  $  {5\times 10^{-10}}\ll {m_{X}}/{M_{\rm pl}}\ll {2\times 10^{-6}} 
\quad   $ &  $  6\times 10^{-13}\lesssim 
{m_Y}/{M_{\rm pl}}\ll  10^{-10}\left(\frac{m_{X}}{M_{\rm pl}}\right)^{1/4}
$ \\ 

$ 4   $  & $   {3\times 10^{-11}}\leq {m_{\chi}}/{M_{\rm pl}}\ll {2\times 10^{-6}}
  $ &  $  
4\times 10^{-14}\lesssim 
{m_Y}/{M_{\rm pl}}\ll  10^{-10}\left(\frac{m_{X}}{M_{\rm pl}}\right)^{1/3}$ \\ 

$ 5   $ & $  {6\times 10^{-12}}\leq {m_{\chi}}/{M_{\rm pl}}\ll {2\times 10^{-6}}$ &  $  
7\times 10^{-15}\lesssim 
{m_Y}/{M_{\rm pl}}\ll  10^{-10}\left(\frac{m_{X}}{M_{\rm pl}}\right)^{3/8}$ \\ 

$ 6   $ & $  {2\times 10^{-12}}\leq {m_{\chi}}/{M_{\rm pl}}\ll {4\times 10^{-7}}$ &  $  
2\times 10^{-15}\lesssim 
{m_Y}/{M_{\rm pl}}\ll  10^{-10}\left(\frac{m_{X}}{M_{\rm pl}}\right)^{2/5}$
\\ 

\hline 
\end{tabular} }
\end{center}
\caption{Range of viable $X$ and $Y$ masses. 
}
\label{table5}

\end{table}
\begin{table}[t]
\begin{center}
\renewcommand{\arraystretch}{1.5}
\resizebox{0.45\textwidth}{!}{
\begin{tabular}{c
@{\hspace{0.5cm}}c}
\hline 
$n$ & $m_{Y}/M_{\rm pl}$ 
\\ 
\hline
\hline

$ 3  $ & $\frac{m_Y}{M_{\rm pl}}\cong 10^{-10}
    \left(  \frac{m_X}{M_{\rm pl}}\right)^{\frac{1}{4}}\Longrightarrow
   7\times 10^{-13}\leq\frac{m_Y}{M_{\rm pl}}\ll 4\times 10^{-12}$  \\ 

$ 4   $ & $\frac{m_Y}{M_{\rm pl}}\cong
     10^{-10}
    \left(  \frac{m_X}{M_{\rm pl}}\right)^{\frac{1}{3}} \Longrightarrow 4\times 10^{-14} \leq\frac{m_Y}{M_{\rm pl}}\ll  10^{-12}$ \\ 

$ 5   $ & $  \frac{m_Y}{M_{\rm pl}}\cong
     10^{-10}
    \left(  \frac{m_X}{M_{\rm pl}}\right)^{\frac{3}{8}} \Longrightarrow    9\times 10^{-15}\leq \frac{m_Y}{M_{\rm pl}}\ll 7\times 10^{-13}$ \\ 

$ 6   $ & $  \frac{m_Y}{M_{\rm pl}}\cong
     10^{-10}
    \left(  \frac{m_X}{M_{\rm pl}}\right)^{\frac{2}{5}}\Longrightarrow
    3\times 10^{-15} \leq\frac{m_Y}{M_{\rm pl}}\ll 5\times 10^{-13}
$ \\ 

$ \infty   $ & $ \frac{m_Y}{M_{\rm pl}}\cong
     10^{-10}
    \left(  \frac{m_X}{M_{\rm pl}}\right)^{\frac{1}{2}}\Longrightarrow
    {4\times 10^{-18}}
    \leq\frac{m_Y}{M_{\rm pl}}\ll  10^{-13}$  \\ 
\hline 
\end{tabular} }
\end{center}
\caption{Range of viable masses for the dark matter. }
\label{table6}
\end{table}

\begin{figure}[ht]
   \centering
\includegraphics[width=0.5\textwidth]{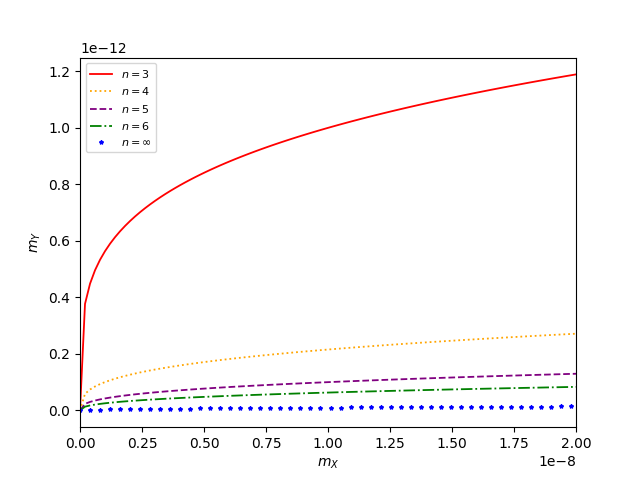}
\caption{Plot of $m_Y$ masses as a function of $m_X$. For:
$n=3$ red, $n=4$ orange, $n=5$ purple,
$n=6$ green and $n=\infty$ blue. In the plot the units are $M_{\rm pl}=1$. 
}
\label{fig:masses}
\end{figure}

{
\subsection{Dark matter production when the reheating temperature is maximum}

As we have already shown, the reheating temperature attains its maximum value when the decay  occurs close to the end of the inflaton's domination. In this situation,  the energy density of the $X$-field at the onset of radiation epoch is $\langle \rho_{X,\rm reh}\rangle=3H_{\rm END}^2M_{\rm pl}^2\Theta_X^{\frac{2n}{n-1}}$ and  the one of the $Y$-field is
$\langle \rho_{Y,\rm reh}\rangle=3H_{\rm END}^2M_{\rm pl}^2\Theta_X^{\frac{n+1}{n-1}}\Theta_Y$, which leads to 
\begin{eqnarray}
    \langle \rho_{Y,\rm eq}\rangle=
3H_{\rm END}^2M_{\rm pl}^2\Theta_X^{-\frac{2(n-2)}{n-1}}\Theta_Y^4.
    \end{eqnarray}
And thus, 
\begin{eqnarray} 
3\frac{H_{\rm END}^2}{M_{\rm pl}^2}\Theta_X^{-\frac{2(n-2)}{n-1}}\Theta_Y^4 \cong 10^{-110}\nonumber\\ \Longrightarrow
12\Theta_X^{-\frac{2(n-2)}{n-1}}\Theta_Y^4\cong 10^{-98}.
\end{eqnarray}
What leads to the following link between both masses 
\begin{eqnarray}\label{X-Y}    \frac{m_Y}{M_{\rm pl}}\cong
    10^{-10}
    \left(  \frac{m_X}{M_{\rm pl}}\right)^{\frac{n-2}{2(n-1)}},
\end{eqnarray}
and thus, as we have already explained, since the mass of the particles responsible for the reheating in relation with the spectral index, one concludes that the mass of dark matter, when one assumes that the onset of radiation is close to the decay of $X$-particles obtaining the maximum reheating temperature,  is related by the spectral index via the formulas 
(\ref{T_max}) and 
(\ref{X-Y}) and (\ref{n_s})

Finally, 
taking into account the range of values obtained for the masses $m_X$ {for moderate values of $n$ (summarized in Table~\ref{table3})
and for large values of $n$ (obtained in section \ref{GW} considering the overproduction of GWs)}, 
we show  in Table \ref{table6} the  range of the values of the masses $m_Y$ for different values of the parameter $n$ and in Figure~ \ref{fig:masses} we show the graphical relation between $m_X$ and $m_Y$ for different values of $n$.

}

}

\section{Conclusions}
\label{sec-summary}

In this article, we examine gravitational reheating as a result of massive particle production when a massive scalar quantum field is conformally coupled to the Ricci scalar. In this framework, calculating the energy density of particles produced at the end of inflation—considering that the number of final 
$e$-folds establishes a relationship between the spectral index of scalar perturbations and the reheating temperature enables us, after obtaining the formula of the reheating temperature,  to link the spectral index to the mass of the generated particles. Assuming a feasible reheating temperature, consistent with Big Bang Nucleosynthesis and avoiding issues related to the gravitino problem, we can place constraints on the possible masses of these particles, and consequently on the spectral index.

Furthermore, we explore the gravitational production of dark matter, deriving relationships between the mass of the particles responsible for reheating and the mass of dark matter particles. These connections provide a natural linkage between the reheating process and the origin of dark matter, offering constraints that enhance our understanding of both early-universe dynamics and dark matter genesis within a gravitational reheating scenario.

Our results indicate that, in the framework of gravitational particle production,  the relationship between the mass of reheating particles and the dark matter mass can yield insights into viable particle mass ranges that comply with observational constraints on the spectral index and reheating temperature. This approach not only strengthens the theoretical basis for gravitational reheating but also aligns with existing cosmological models that incorporate  gravitational dark matter production. Thus, by establishing mass bounds, this study contributes to a more precise picture of the mechanisms at play during the transition from inflation to the reheated universe, connecting fundamental aspects of inflationary theory with dark matter production pathways.

\section*{Acknowledgments} 
{The authors thank the referee for some important comments that improved the quality of the manuscript.} JdH is supported by the Spanish grants 
PID2021-123903NB-I00 and 
RED2022-134784-T
funded by MCIN / AEI / 10.13039/501100011033 and ERDF ``A way of making Europe''.
SP has been supported by the Department of Science and Technology (DST), Govt. of India under the Scheme   ``Fund for Improvement of S\&T Infrastructure (FIST)'' (File No. SR/FST/MS-I/2019/41).   

\bibliography{references}

\end{document}